\documentclass[12pt]{article}
\usepackage{subfigure}
\usepackage{amssymb,amsmath}
\usepackage{graphicx}
\usepackage{float}
\usepackage{color}
\usepackage{array,multirow}
\usepackage[colorlinks=true
,urlcolor=blue
,citecolor=blue
,linkcolor=blue
,pagecolor=blue
,linktocpage=true
,pdfproducer=medialab
]{hyperref}
\usepackage[a4paper]{geometry}
\usepackage{cite}
\usepackage{placeins}
\usepackage[utf8]{inputenc}
\usepackage{cancel}
\usepackage{arydshln}
\makeatletter \renewcommand{\@dotsep}{10000} \makeatother
\usepackage{indentfirst}
\def\beq{\begin{equation}}
\def\eeq{\end{equation}}
\usepackage[nodisplayskipstretch]{setspace}

\begin{document}

\begin{titlepage}

\begin{flushright}

\end{flushright}
\pagestyle{empty}

\vspace*{0.2in}
\begin{center}
{\Large \bf An Information Theoretic Exploration of Constrained \\[0.25cm] MSSM}\\
\vspace{1cm}
{\bf  Surabhi Gupta$^{a}$\footnote{E-mail:sgupta2@myamu.ac.in } and Sudhir Kumar Gupta$^{a}$\footnote{E-mail:sudhir.ph@amu.ac.in} }
\vspace{0.5cm}
\begin{flushleft}
{\it $^a$Department of Physics, Aligarh Muslim University, Aligarh, 
UP--202002, INDIA} \\

\end{flushleft}

\end{center}

\vspace{0.5cm}
\begin{abstract}

\end{abstract}

 We discuss information theory as a tool to investigate constrained minimal supersymmetric Standard Model (CMSSM) in the light of observation of Higgs boson at the Large Hadron Collider. The entropy of the Higgs boson using its various detection modes has been constructed as a measure of the information and has been utilized to explore a wide range of CMSSM parameter space after including various experimental constraints from the LEP data, B-physics, electroweak precision observables and relic density of dark matter. According to our study while the lightest neutralino is preferred to have a mass around 1.92 TeV, the gluino mass is estimated to be around 7.44 TeV. The values of CMSSM parameters $m_0$, $m_{1/2}$, $A_0$ and $\tan\beta$ correspond to the most preferred scenario are found to be about 6 TeV, 3.6 TeV, $-$6.9 TeV and 36.8 respectively.

\end{titlepage}

\section{Introduction}
\label{sec:intro}

The recent discovery of Higgs boson and the subsequent measurement of its mass through various detection modes 
at the Large Hadron Collider (LHC)~\cite{Aad:2015zhl} have been found to be consistent with the Standard Model 
(SM) predictions. However the SM~\cite{Djouadi:2005gi}, due to lack of providing a satisfactory stability 
mechanism which could prevent its mass to grow quadratically up the Planck scale, ambitions to realize the 
existence of dark matter, grand unification, finite values of masses of neutrinos, and lack of gravitational 
interactions, itself provides room to investigate theories beyond the SM. Among several interesting candidate 
theories, supersymmetry (SUSY)~\cite{Martin:1997ns, Tata:1997uf, Drees:1996ca, Aitchison:2005cf, Fayet:2015sra, 
Djouadi:2005, Cane:2019ac, Allanchach:2019wrx} is still sought to be one of the most preferred choices as it is 
capable of solving most of the aforementioned problems. It also opens up the scope of looking for counterparts 
of the SM particles known as the superparticles which are differing by spin-1/2 in the minimal extension of SM, 
known as the minimal supersymmetric Standard Model (MSSM)~\cite{Aitchison:2005cf, Djouadi:2005, 
Tanabashi:2018oca, Dawson:1997tz, Heinemeyer:1998np, Draper:2016pys}. Besides it, the Higgs sector in MSSM is 
enhanced by one CP-even heavier neutral Higgs boson ($H$), a CP-odd neutral Higgs boson ($A^0$) and charged Higgs 
boson ($H^\pm$). The lack of any experimental evidence in support of SUSY hint that if it exists, it must have 
been broken at a scale much higher than the electroweak symmetry breaking (EWSB) scale which is of about 100 GeV 
or so. The aim of this study is therefore to understand the SUSY breaking scale which is preferred by the Higgs 
mass observation together with the data from other experiments. Although there have been a lot of efforts in 
this direction which is either driven by the frequentist~\cite{Ellis:2012aa,Buchmueller:2011ki,Bechtle:2012zk} 
or by the Bayesian~\cite{Fowlie:2014xha,Fowlie:2011mb,Allanach:2011wi} framework. In our article we employ a 
completely new approach to the SUSY searches which relies upon the information theory. Information theory which 
is primarily based upon Shannon's entropy has already been applied successfully in studying particle 
physics as well as in several areas of physics including cosmology to yield remarkable 
results~\cite{Alves:2014ksa,Ensslin:2008iu, Pinho:2020uzv, Pandey:2015reu, Bernardini:2016hvx, Alves:2020cmr, Alves:2017ljt, dEnterria:2012eip, Millan:2018fme, Jizba:2019cac, 
Wang:2014lua, Hamada:2014xra, Llanes-Estrada:2017clj, B:2018thi, Nosek:2015mse, Fang:2015pcw, Quznet:2020}. We find that the first work on precise estimation of SM Higgs mass which is close to experimental value through maximization of the product of its branching ratios in Ref.\cite{dEnterria:2012eip}. In Ref.\cite{Alves:2014ksa} the entropy of the Higgs is constructed using the branching ratios of the SM Higgs boson through its various decay modes at the LHC. 
According to the finding of this article the maximum entropy technique is capable of predicting the Higgs mass very close to its measured value at the LHC. The maximum entropy principle has also been successfully applied in the context of studying new decay modes of Higgs boson at the LHC in Ref.\cite{Alves:2020cmr}. Furthermore Ref.\cite{Alves:2017ljt} suggests that this technique is also capable of estimating axion mass through axion-neutrino interaction in effective field-theoretic models. In~\cite{Millan:2018fme,Llanes-Estrada:2017clj}, Gibbs-Shannon entropy has been incorporated for the distribution of the decay channels of the hadrons and measure the information of the newly added decay channel to it.

The organization of the article is as follows: In Section~\ref{sec:entropy} we discuss information theory in the context of LHC Higgs observations. The investigation on Higgs boson in the CMSSM and other sparticle masses using information theory as a tool is discussed in Section~\ref{sec:CMSSM}. Finally we discuss our findings in Section~\ref{sec:result}.

\section{Information Theory and Higgs Observation}
\label{sec:entropy}

The information theory provides a quantitative measure of uncertainty in finding the state of a system using an analog of Boltzmann entropy which is known as the Gibbs-Shannon entropy or information entropy, i.e. the lack of information, randomness or a disorder translate into the rising size of the entropy. The Boltzmann entropy $S$ of a thermodynamic system is defined for all possible microstates of the system under consideration via $S = k_B \ln{\Omega}$, with $k_B$ being the Boltzmann's constant and $\Omega$ represents the number of microstates associated with the macrostate of the system. For the system having all the microstates with universal probability $p = 1/\Omega$, the above equation would translate into the following form 
\beq
\label{eq:y:0}
S = - k_B \ln{p}. 
\eeq
The aforementioned equation indicates that smaller the probability of microstates in the same macroscopic system, the larger would be the entropy of the system.\\
As the entropy is an additive measure, in case the microstates differ in their probabilities, the above equation could be modified by extending entropy to its expectation value, i.e. for the $i^{th}$ microstate having probability $p_i$, the entropy of the system would then be
\beq\label{eq:y:1} 
S =  - k_B \left<\ln{p}\right> = -k_B \sum_{i\in\{micro\}} p_{i} \ln{p_{i}}. 
\eeq
Shannon~\cite{shannon,jaynes:1957,thomas:2006} considered entropy to be a measure of uncertainty of the information content and defined the information entropy (or Shannon's entropy) like Eq.~\eqref{eq:y:1} with $k_B = 1$. Thus the information theory is basically the probability theory where probabilities represent our ignorance of an event and the probability distribution contains information of every event then the amount of information or entropy is associated with the expectation value of the random variable of the event. The negative of the logarithm of the probability distribution is considered to be information. The probabilities of the probability distribution lie from zero to one, the total of all is unity as the events of this distribution are mutually independent and exhaustive. These probabilities are required only when none of the probabilities is equal to one i.e. there is uncertainty in the state of the system. 

To make it more clear consider the example of a coin toss with two possible outcomes head and tail have an equal probability. This yields the information entropy, $S$ = $-$(1/2) $\ln$({1/2}) $-$(1/2) $\ln ({1/2}) = \ln {2}$ (or 0.693) nats of information, nat is a unit of entropy. If a coin has two heads and no tail or two tails and no head then entropy would be zero (no uncertainty). Here the outcome is previously known so this result gives no new information. If the probability of one side of a coin is more than the other side, the outcome gives reduced uncertainty. This implies lower entropy which comes out to be less than $\ln {2}$ nats of information. Rarer events give more information than the probable events. Or in other words, maximum entropy means a stable configuration  as well as maximum ignorance of the system. This is known as the principle of maximum entropy and is quite essential to determine the probability distribution corresponds to the maximum value for the uncertainty, without having any previous information. Thus using the principle of maximum entropy, a value corresponding to the variable of the probability distribution can be determined effectively.  

\begin{figure}
\begin{centering}
\includegraphics[angle=0,width=0.7\linewidth,height=16.4em]{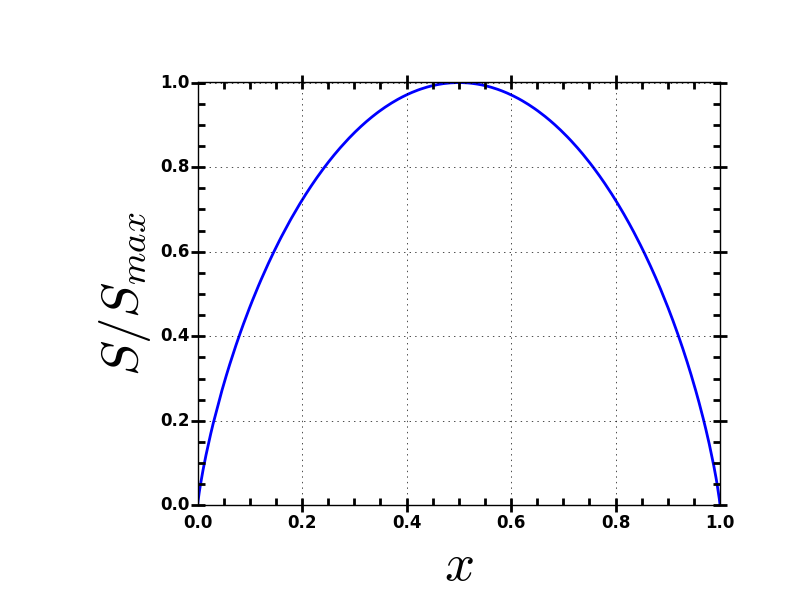}
\caption{\sf{Entropy for a coin toss vs the success probability of obtaining head or tail $x$.}}
\label{fig:coin}
\end{centering} 
\end{figure}

Assuming $x \in [0, 1]$ to be success probability of obtaining a head $(h)$ in coin toss, i.e. $p_1 \equiv p(h) = x$ and $p_2 \equiv p(t) = 1 - x$, the entropy $S(x)$ could be written as
\beq
\label{eq:y:2}
S(x) = -  p_1 \ln{p_1} - p_2 \ln{p_2} = - (x) \ln({x}) - (1 - x) \ln{(1-x)},
\eeq
with the condition that the entropy maximizes at $x = x_{max}$, i.e. 
\beq
\label{eq:y:3} 
\dfrac{d S(x)}{dx}\rvert_{_{x_{max}}} = 0. \\
\eeq
From the above expression it is straightforward to note that for $x_{max} = 0.5$, both $h$ and $t$ are equally probable and entropy turns out to be maximum with a value equal to $\ln 2$ or $0.693$ nats of information. A distribution of $S/S_{max}$ vs $x$ is shown in Figure~\ref{fig:coin}. As we move away from the $x_{max}$ the information content is reduced due to the rising imbalance between the probabilities of $h$ or $t$. 

\begin{figure}
\includegraphics[angle=0,width=0.9\linewidth,height=16em]{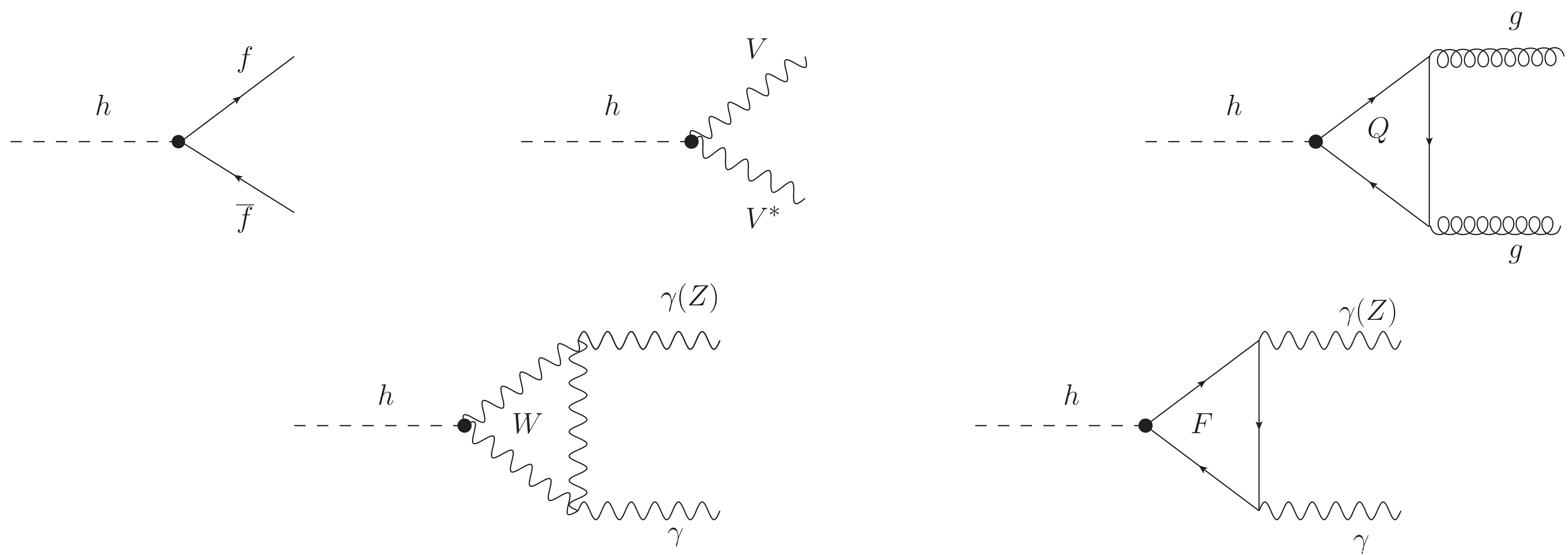}
\caption{\sf{Representative Feynman diagrams for the SM Higgs decays, $f \in$ [$u, d, c, s, b, e^\pm, \mu^\pm, \tau^\pm$] and $V \in$ [$W^\pm, Z$].}}
\label{fig:smhiggsfym}
\end{figure}

\begin{figure}
\begin{centering}
\includegraphics[angle=0,width=0.60\linewidth,height=16em]{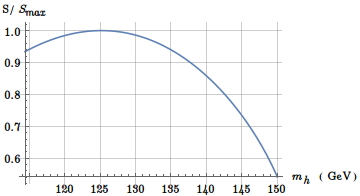}
\caption{\sf{Marginalized entropy vs Higgs boson mass for the SM.}}
\label{fig:smhiggs}
\end{centering}
\end{figure}

The above idea is directly applicable to study new physics models through the observed Higgs boson at the LHC by assuming $\cal N$-number of independent Higgs bosons forming an ensemble wherein each of them is allowed to decay through its various decay modes permissible by the underlying new physics theory with probabilities $p_{_d} (m_h)$ given by the branching ratio $Br_d (m_h)$ of the respective decay channel, i.e.  
\beq
\label{eq:y:4} 
 p_{_d} (m_h) \equiv Br_d (m_h) =  \frac{\Gamma_d (m_h)}{\Gamma_h (m_h)}, 
 \eeq
where $\Gamma_d (m_h)$ represents the partial decay width of Higgs boson in its $d^{th}$ decay mode and $\Gamma_h (m_h) = \sum_{d = 1}^{n_d}\Gamma_{d}(m_h)$ being the total decay width of the Higgs boson, $n_d$ to be the  number of allowed decay modes of the Higgs boson.
The probability that this  ensemble reaches to a final state through its various decay modes is given by the following  multinomial distribution~\cite{multi} as mentioned in~\cite{Alves:2014ksa} 
\beq
\label{eq:y:5}
{\cal P}_{\{m{_d}\}}(m_h) = \frac{{\cal N}!}{m_1!...m_{n_d}!}\prod_{d = 1}^{n_d}{(p_{_d} (m_h))}^{m_d},
\eeq
with $\sum_{d = 1}^{n_d}{Br}_{d} = 1 $  and $ \sum_{d = 1}^{n_d}m_{d} = {\cal N}$; $m_{d}$ represents the number of Higgs bosons decaying through $d^{th}$ detection mode. The total information entropy associated with the ${\cal N}$-Higgs bosons in their final state, as specified in~\cite{Alves:2014ksa}, could therefore be given by the following formula 
\beq
\label{eq:y:6} 
S (m_h) = - \sum^{\cal N}_{\lbrace m_d \rbrace} {\cal P}_{\{m{_d}\}}(m_h) \ln {\cal P}_{\{m{_d}\}}(m_h).
\eeq
An asymptotic expansion~\cite{cicho:hal} of the above expression would result in the following form of entropy of the Higgs boson~\cite{Alves:2014ksa}, 
\beq
\label{eq:y:7}
S (m_h) \simeq \frac{1}{2}\ln\left(\left(2\pi {\cal N} e\right)^{n_d -  1} \prod_{d = 1}^{n_d}{p_{_d} (m_h)}\right) + \frac{1}{12 {\cal N}}\left( 3 n_d - 2 - \sum_{d = 1}^{n_d}{(p_{_d} (m_h))}^{-1}\right)+ {\cal O}\left({\cal N}^{-2}\right).
\eeq
The above equation suggests that in order to get a clear idea of which Higgs mass maximizes the entropy, it is very important to have precise information about the branching ratios of the Higgs as a function of mass as suggested in~\cite{dEnterria:2012eip} and in agreement with the study of~\cite{dEnterria:2012eip}. We therefore use the package {\tt HDECAY}~\cite{Djouadi:2018xqq} to estimate various branching ratios at the NNLO + NLL wherever possible. To demonstrate that the method indeed yields a Higgs mass as observed at the LHC, we consider the SM Higgs boson with a mass range $m_h \in [114.4, 150]$. For this mass range the possible kinematically accessible detection modes of the Higgs are as follows: $h\rightarrow\gamma\gamma$, $h\rightarrow \gamma Z$, $h\rightarrow Z Z^*$, $h\rightarrow W W^*$, $h\rightarrow gg$, $h\rightarrow f\bar{f}$ with $f\in \{u, d, c, s, b, e^\pm, \mu^\pm, \tau^\pm\}$. The leading order Feynman diagrams representing each of these decay channels are displayed as Figure~\ref{fig:smhiggsfym}. Notice that while the fermionic decays of the Higgs boson occur directly and are proportional to the respective Yukawa coupling square, photonic and $h\to \gamma Z$ decays of the Higgs boson take place via loops mediated by charged fermions and $W^\pm$ and thereby highly suppressed. While the gluonic decay which takes place indirectly via the mediation of only quarks loops is also quite suppressed~\cite{Djouadi:2005gi}. As the $S(m_h)$ also depends on other SM parameters such as the fermion and gauge boson masses, coupling strengths and other parameters, values to each of which are already known through earlier experiments, we marginalized $S(m_h)$ with respect to these SM parameters over their experimentally measured values as listed in Table~\ref{tab:table1} for each value of $m_h$. A distribution showing the marginalized entropy, $S$, scaled by a normalization factor $1/S_{max}$ vs the Higgs boson mass, $m_h$, is displayed in Figure~\ref{fig:smhiggs}. According to this distribution the maximum value of entropy corresponds to $m_h \simeq 125.2 \pm 3.3$ GeV.

\begin{table}[h]
 \begin{center}
 \begin{tabular}{ll}
    \hline
    \hline
\textbf{Parameter}  & \textbf{Experimental Value}\\
       \hline  
$ \alpha_s(m_Z)$ & 0.1181   \\
$ sin^2\theta_W$ & 0.22343  \\
$ G_F$ & 1.1663787 $\times$ $10^{-5}$ GeV$^{-2}$\\
$ m_b(m_b)$ & 4.18 GeV \\
$ m_W$ &  80.379 GeV \\
$ m_Z$ &  91.1876 GeV \\
$ m_t(m_t)$ & 173.1 $\pm$ 0.9 GeV\\
$ m_\tau$ & 1.7768 GeV 
 \\
   \hline
   \hline
    \end{tabular}
    \caption{\sf{The SM parameters and their experimental values used in our analysis}~\cite{Tanabashi:2018oca}.}
    \label{tab:table1}
   \end{center}
\end{table}

\section{CMSSM and Information Theory}
\label{sec:CMSSM}

As discussed before MSSM was introduced to stabilize the Higgs mass against the radiative corrections by relating the bosons and the fermions with each other. This requires an equal number of bosonic and fermionic degrees of freedom, i.e. each SM particle will have corresponding SUSY counterpart which differs by spin-1/2. In terms of `R-parity`, $R = (-1)^{L + 3B + 2S}$ where $S$, $L$ and $B$ represent spin, lepton number and baryon number respectively, one can infer that each R-even SM particle will have an R-odd SUSY partner. Assuming R-parity to be conserved, the MSSM is specified by the superpotential~\cite{Tanabashi:2018oca}
\beq
\label{eq:y:8}
V_{MSSM} = Y_d \hat{H}_d \hat{Q} \hat{D}^c - Y_u \hat{H}_u \hat{Q} \hat{U}^c + Y_l \hat{H}_d \hat{L} \hat{E}^c + \mu \hat{H}_u \hat{H}_d,
\eeq
where, $ \hat{Q},\hat{L}, \hat{D}^c, \hat{U}^c, \hat{E}^c$, $\hat{H}_u$ and $\hat{H}_d$ represent left-handed doublet quark superfield, left-handed doublet lepton superfield, right-handed down-type anti-quark singlet superfield, right-handed up-type anti-quark singlet superfield, right-handed anti-lepton singlet superfield, up-type and down-type Higgs doublet chiral superfields respectively and $Y_l$, $Y_d$ and $Y_u$ are Yukawa matrices for leptons, down- and up-type quarks respectively. The Bilinear term represents the mixing of up- and down-type Higgs doublets which are required to provide masses to the up- and down-type sfermions separately. 

In this paper we analyze Higgs boson for the constrained MSSM (CMSSM)~\cite{Athron:2017fxj, Balazs:2013qva, Fowlie:2012im, allenach, Kane:1993td} which is most popular due to its aesthetic built-up in terms of only five parameters at the GUT scale, namely the universal scalar and gaugino masses $m_0$ and $m_{1/2}$ respectively, a common trilinear coupling parameter $A_{0}$, the ratio of vacuum expectation values (VEVs) of up- and down-type Higgs bosons $\tan\beta$ and the sign of the Higgsino mass parameter $\mu$, $sign(\mu)$. The mass spectrum of CMSSM could be obtained by running the renormalization group equations (RGEs) down to EW scale from GUT scale which turns out to be different due to individual RGE evolution patterns responsible for each of these. In the CMSSM, the tree-level mass of the lighter CP-even Higgs boson is given by 
\beq
\label{eq:y:9}
 m^{2}_{h} = m_{Z}^2\cos^{2}2\beta, 
 \eeq
and therefore $m_h \lesssim m_Z$ at the tree-level. This is corrected by the following leading order correction~\cite{Tanabashi:2018oca}
\beq
\label{eq:y:10}
\triangle m^2_{h} = \dfrac{3 g^2 m_t^4}{8\pi^2 m_W^2}[ln(\dfrac{M_S^2}{m_t^2})+ \dfrac{X_t^2}{M_S^2}(1-\dfrac{X^2_t}{12M_S^2})],
\eeq
where $M_S^2 \equiv m_{\tilde t_1} m_{\tilde t_2}$, $m_{\tilde t_{1,2}}$ are the masses of the superpartners of top-quark, the stop mixing parameter $X_t = A_t - \mu cot\beta$ which arises due to an incomplete cancellation of the top-quark and top-squark loops getting different masses in supersymmetry breaking. This suggests that the correction in the Higgs mass depends on a factor of $m_t^4$ and grows logarithmically with the stop mass which ensures that the Higgs mass still remains well within $150$ GeV or so even for the SUSY breaking scale in the multi-TeV range.

Now since for TeV scale SUSY breaking all the sparticles including the lightest stable particle (LSP) which is usually the lightest neutralino is also heavier than the lighter CP-even Higgs boson, other than the SM decay modes, no further decay channels open up for the Higgs boson. However the presence of sparticles in the loops modify these decay channels via higher-order contribution. The decay channels involving loops at the leading order such as $h\to \gamma\gamma$ and $h\to Z\gamma$ will even receive significant contributions at the first order itself~\cite{Djouadi:2005}.

We therefore analyze the possibility whether the information theory which solely requires the information about the branching ratios of the Higgs boson could shed light on the Higgs mass pretending it to be unknown and thereby guide us about the other sparticle masses simultaneously. These findings are further improved by imposing further constraints from the LEP data on sparticles and Higgs searches, experimental data on electroweak precision observables (EWPOs) such as the muon anomalous magnetic moment $a_\mu$ and contribution to $\rho$ parameter, branching ratios of $b \to s \gamma$ and $B^0_s \to \mu^+\mu^-$ and relic density of dark matter $\Omega_{\chi}h^{2}$. Bounds corresponding to each of these are listed in Table~\ref{tab:table2}. With this in mind, we perform a detailed parameter scan of the CMSSM with the ranges of various CMSSM parameters specified below
\begin{itemize}
\item $m_0 \in [0.1, 10] $ TeV,
\item $m_{1/2} \in [0.1, 6] $ TeV,
\item $tan\beta \in [2, 60]$,
\item $A_{0} \in [-10, 10]$ TeV,
\item sign$(\mu)$ = +1.
\end{itemize}

In order to generate CMSSM spectrum we use {\tt Softsusy 4.1.3}~\cite{Allanach:2001kg}. This is then interfaced with {\tt FeynHiggs 2.14.2}~\cite{Hahn:2013ria, Bahl:2018qog} to calculate the branching ratios of Higgs boson and $\rho$ parameter, {\tt Superiso v4.0}~\cite{Arbey:2018msw} for estimating muon anomalous magnetic moment and B-physics branching ratios and {\tt micromegas 
5.0.4}~\cite{Belanger:2004yn, Belanger:2001fz} to calculate the relic density of the dark matter which in our case turn out to be lightest neutralino.

\begin{table}[h]
 \begin{centering}
    \tabcolsep 0.4pt
    \small
    \begin{tabular}{ccc}
    \hline
    \hline
    \textbf{Constraints}&\textbf{Observables}  & \textbf{ Experimental~Values}\\
       \hline 
LEP &$ m_{h}$ &  $>$ 114.4 GeV \cite{Schael:2006cr}\\
&$ m_{\tilde\chi^{0}_{1,2,3,4}}$ & $>$ 0.5 $m_Z$ \cite{Tanabashi:2018oca}\\
&$ m_{\tilde\chi^{\pm}_{1,2}}$& $>$ 103.5 GeV \cite{Tanabashi:2018oca}\\
\hline
PO & $ \triangle\rho$  & $0.0008\pm0.0017$ \cite{Nakamura:2010zzi}\\
&$ BR(b \to s\gamma)$ & $(3.55\pm0.24)\times10^{-4}$ \cite{Asner:2010qj}\\
&$BR(B^0_s \to \mu^+\mu^-)$& $(3.0\pm0.6)\times10^{-9}$ \cite{Aaij:2017vad}\\
& $\triangle a_{\mu}$ & $(2.68\pm0.43)\times10^{-9}$ \cite{Tanabashi:2018oca}\\
\hline
DM & $ \Omega_{\chi}h^{2}$ & $0.1186\pm0.002$ \cite{Tanabashi:2018oca}\\
   \hline
    \hline
    \end{tabular}
     \caption{\sf{Constraints of various experimental observables.}}
      \label{tab:table2}
   \end{centering}
\end{table}

\begin{figure}
\begin{centering}
\includegraphics[angle=0,width=0.65\linewidth,height=19em]{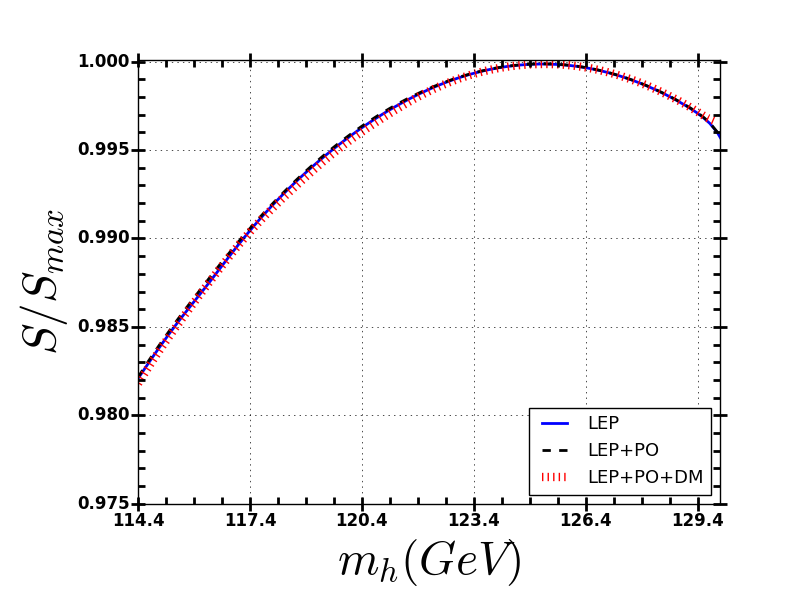}
\caption{\sf{Marginalized entropy vs CP-even lighter Higgs mass. The blue solid line represents only LEP constraints, the black dashed line represents LEP and PO constraints and the red dotted line represents LEP, PO and DM constraints. The details of constraints are listed in Table~\ref{tab:table2}.}}
\label{fig:MH}
\end{centering} 
\end{figure}

\begin{figure}
\includegraphics[angle=0,width=0.329\linewidth,height=16.5em]{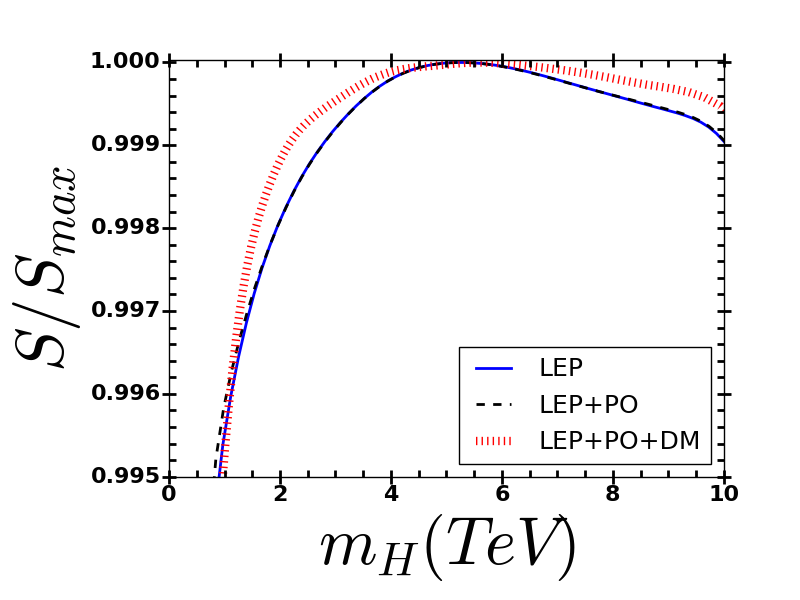}
\includegraphics[angle=0,width=0.329\linewidth,height=16.5em]{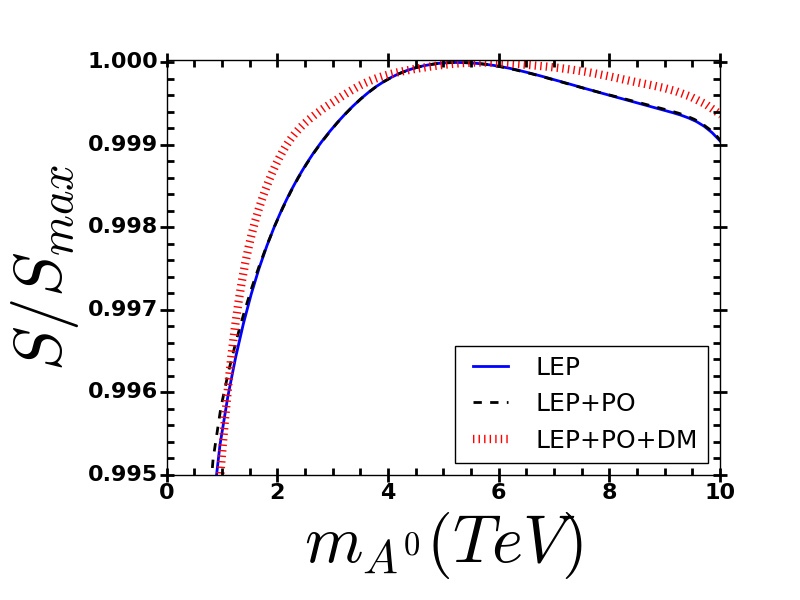}
\includegraphics[angle=0,width=0.329\linewidth,height=16.5em]{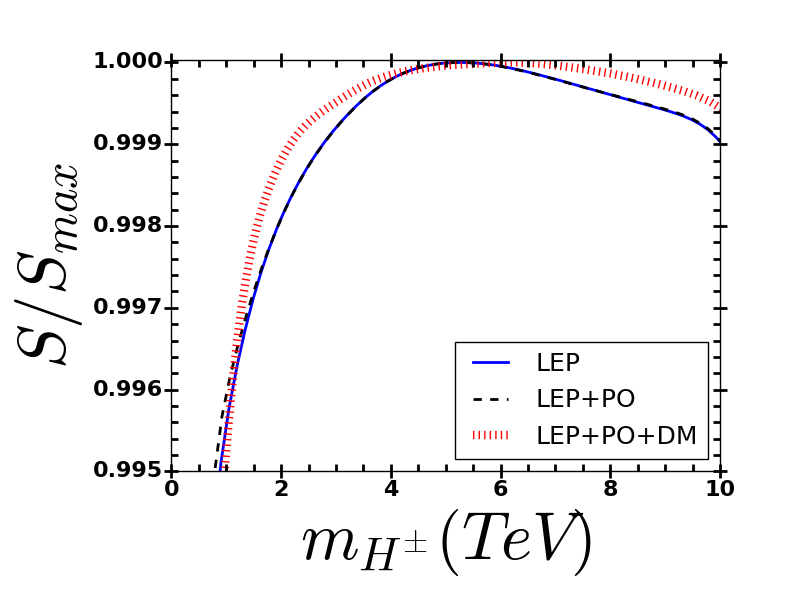}
\caption{\sf{Marginalized entropy vs mass of the (a) heavier CP-even neutral Higgs boson $H$ (left), (b) CP-odd neutral Higgs boson $A^0$ (middle) and (c) charged Higgs boson $H^\pm$ (right). Color convention is the same as in Figure~\ref{fig:MH}.}}
\label{fig:MH1}
\end{figure}

\begin{figure}
\includegraphics[angle=0,width=1.12\linewidth,height=61em]{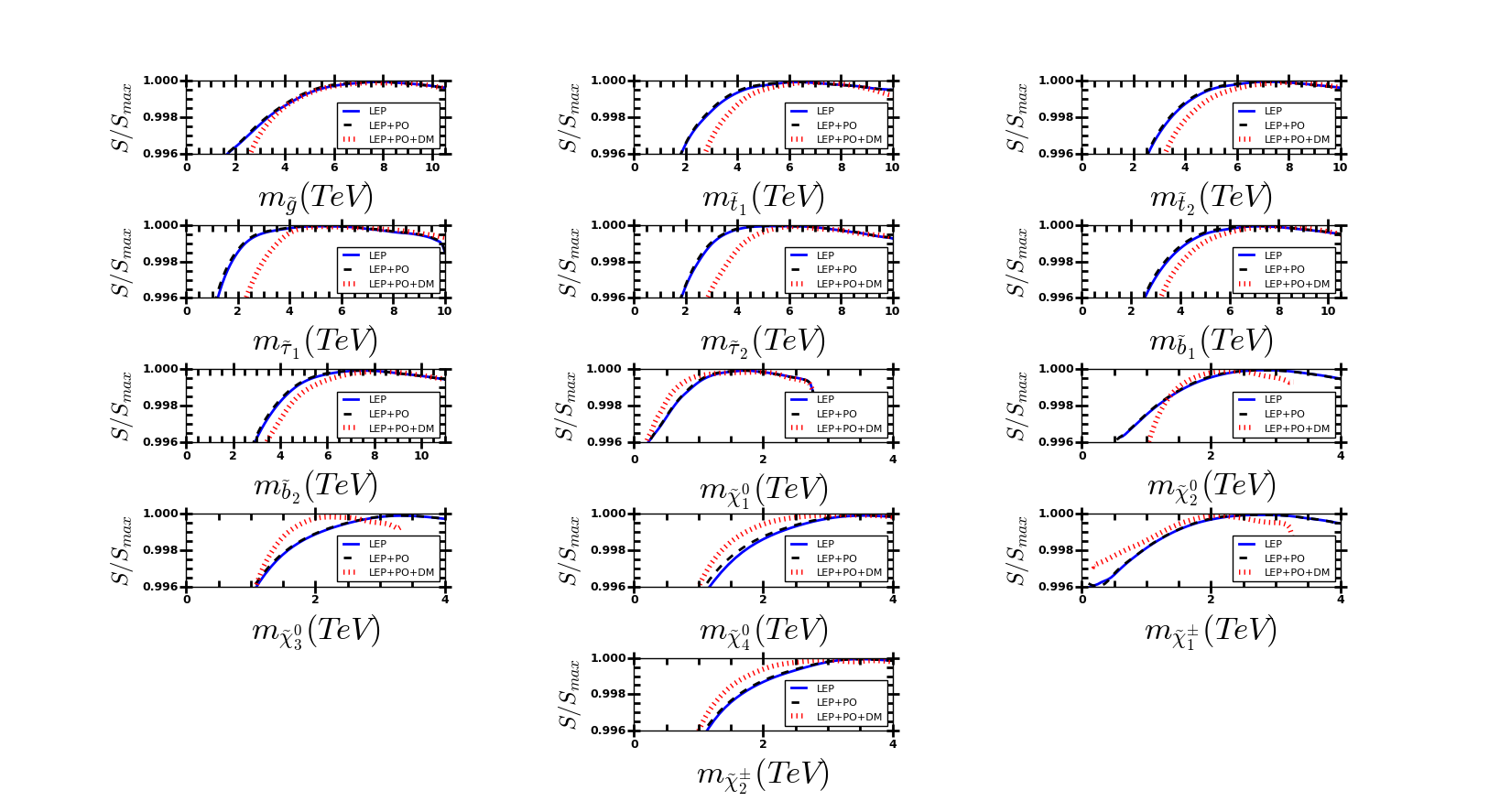}
\caption{\sf{Marginalized entropy vs sparticle masses. Color convention is the same as in Figure~\ref{fig:MH}.}}
\label{fig:OUT}
\end{figure}

\begin{figure}
\includegraphics[angle=0,width=1.1\linewidth,height=25.5em]{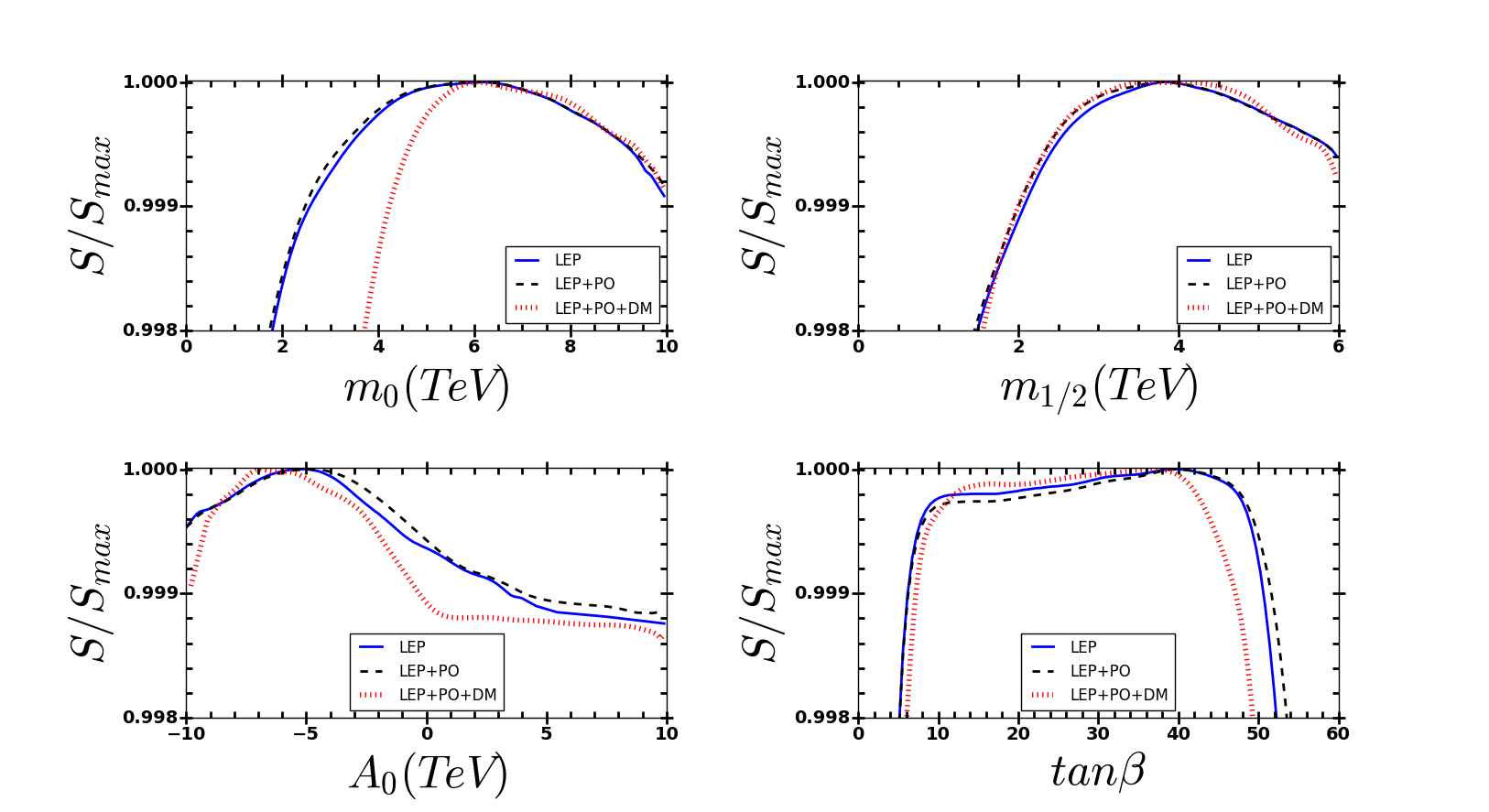}
\caption{\sf{Marginalized entropy vs CMSSM parameters. Color convention is the same as in Figure~\ref{fig:MH}.}}
\label{fig:IN}
\end{figure}

An entropy measure for an ensemble of the CP-even lighter Higgs boson in the CMSSM using Eq.~\eqref{eq:y:7} follows the same analytic procedure as discussed in Section~\ref{sec:entropy}, i.e. we first calculate the kinematically accessible detection modes of the CP-even lighter Higgs boson are as follows: $h\rightarrow\gamma\gamma$, 
$h\rightarrow \gamma Z$, $h\rightarrow Z Z^*$, $h\rightarrow W W^*$, $h\rightarrow gg$, $h\rightarrow f\bar{f}$ with $f\in \{u, d, c, s, b, e^\pm, \mu^\pm, \tau^\pm\}$. {\tt FeynHiggs 2.14.2}~\cite{Hahn:2013ria, Bahl:2018qog} is used for calculating the Higgs masses and their branching ratios. Based on the information theoretic approach, the entropy associated with the Higgs boson decays to all available decay channels is maximized for a given Higgs boson mass. $S$ is then marginalized entropy having a function of $m_h$ only, obtained after marginalizing over all other parameters of the MSSM and scaled by a normalization factor $1/S_{max}$. We use various constraints in our phenomenological analysis which are based on experimental values as given in Table~\ref{tab:table2}. The restrictions due to LEP data on Higgs mass, neutralino and chargino masses are ensured by imposing $ m_{h} >$ 114.4 GeV, $ m_{\tilde\chi^{0}_{1,2,3,4}} >$ 0.5 $m_Z$ and $ m_{\tilde\chi^{\pm}_{1,2}} >$ 103.5 GeV respectively whereas the bounds on $ \bigtriangleup\rho$, $\triangle a_{\mu}$, $ BR(b \to s\gamma)$, $ BR(B^0_s \to \mu^+\mu^-)$ and bound on relic density of dark matter $\Omega_{\chi}h^2$ as listed in Table~\ref{tab:table2} are employed at 2.5$\sigma$ confidence level.
We present our results for three different combinations of constraints namely, (i) LEP only, (ii) LEP $+$ PO and (iii) LEP $+$ PO $+$ DM. These results are presented as Figures~\ref{fig:MH}--\ref{fig:IN}.

In Figure~\ref{fig:MH} we show the variation of marginalized entropy with the CP-even lighter Higgs mass $m_h$ which clearly exhibits similar trend and shows that the entropy maximizes for $m_h \simeq 125.2$. The exact values corresponding to the maximum entropy for all three curves are listed in Table~\ref{tab:table3} and are found to be in good agreement with the experimentally observed value at the LHC ($ m_h = $ 125.09 $\pm$ 0.24 GeV, combined ATLAS and CMS measurements)~\cite{Aad:2015zhl}. It is to be noted here that the information theoretic interpretation of the rapid fall of entropy is associated with the transfer of the Higgs mass to its dominant decay channel. Similar plots for the other Higgses are presented as Figure~\ref {fig:MH1} which suggest that these should be found around 5 TeV or so. In a similar manner we also plot entropy vs various sparticle masses in Figure~\ref {fig:OUT} and listed the values favorable to our approach in Table~\ref{tab:table3}. These plots suggest that the maximum entropy corresponds to a gluino and lighter stop of about 7.44 TeV and 6.75 TeV respectively after including all the constraints. Similarly the most preferred value of masses of lightest neutralino and lighter chargino are 1.92 TeV and 2.1 TeV respectively according to these plots. Finally to see what values of CMSSM parameters correspond to, we present the Higgs entropy for each of these as Figure~\ref {fig:IN}. These suggest $m_0$, $m_{1/2}$, $A_0$ and $\tan\beta$ to be about 6.16 TeV, 3.80 TeV, $-$5.40 TeV and 39.4 respectively with LEP constraints only. These changes to 6.00 TeV, 3.74 TeV, $-$4.93 TeV and 39.6 after including EWPOs and B-Physics constraints, and 5.99 TeV, 3.58 TeV, $-$6.92 TeV and 36.8 once the dark matter relic density constraint is imposed.

\begin{table}[h!]
  \begin{center}
    \small
    \begin{tabular}{lccc} 
      \hline
      \hline
      \multirow{2}{*}{\textbf{Parameters}}& \multicolumn{2}{r}{\textbf{Constraints}}& \\
      \cline{2-4} 
      & \textbf{LEP} & \textbf{LEP $+$ PO} & \textbf{LEP $+$ PO $+$ DM}\\
       \hline
      $ m_{0}$  &6.16&6.00&5.99\\
      $ m_{1/2}$  &3.80 &3.74 &3.58\\
      $ A_{0}$  &$-$5.40 &$-$4.93 &$-$6.92\\
      $ tan\beta$  & 39.4&39.6 &36.8\\
      \hline
      $m_h$  &125.16  & 125.16 & 125.24 \\
      \hline
      $m_{H}$  & 5.21&5.22&5.61\\
      $m_{A^0}$  & 5.20&5.19&5.71\\
      $m_{H^{\pm}}$  &5.27 &5.28&6.19\\
      $ m_{\tilde\chi^{0}_{1}}$  &1.73&1.71&1.92\\
      $ m_{\tilde\chi^{0}_{2}}$  &2.84&2.82&2.32\\
      $ m_{\tilde\chi^{0}_{3}}$   &3.31&3.35&2.23 \\
      $ m_{\tilde\chi^{0}_{4}}$   & 3.55&3.50&3.11\\
      $ m_{\tilde\chi^{\pm}_{1}}$   &2.68 &2.71&2.10\\
      $ m_{\tilde\chi^{\pm}_{2}}$   & 3.45&3.45&2.96\\
      $ m_{\tilde{g}}$  &7.78&7.73&7.44\\
      $ m_{\tilde{q}_L}$  &8.17&7.99&8.97\\
      $ m_{\tilde{q}_R}$  &7.81&7.76&8.71\\
      $ m_{\tilde{b}_1}$   &7.10 &7.09&7.94\\
      $ m_{\tilde{b}_2}$  &7.50&7.49&8.33\\
      $ m_{\tilde{t}_1}$  & 6.24&6.24&6.75\\
      $ m_{\tilde{t}_2}$  & 7.15&7.12&7.96\\
      $ m_{\tilde{l}_L}$  &5.73&5.73&6.60\\
      $ m_{\tilde{l}_R}$  &6.06&5.76&6.18\\
      $ m_{\tilde{\tau}_1}$  &5.31&5.12&5.17\\
       $ m_{\tilde{\tau}_2}$  &5.45&5.35&6.09 \\
      \hline
      \hline
    \end{tabular}
    \caption{\sf{The sparticle mass spectrum for the marginalized maximum entropy for various experimental constraints as listed in Table~\ref{tab:table2}. All parameters except $ tan\beta$ with mass dimension, $m_{h}$ is in GeV and rest sparticles are in TeV.}}
    \label{tab:table3}
 \end{center}
\end{table}


\section{Results and Discussion}
\label{sec:result}

In this work we investigated the CMSSM parameter space using information theory in the light of the recent discovery of a Higgs boson at the LHC. To do so we first estimated entropy of the Higgs boson through its various decay modes. Later we used it to check the mass of the Higgs boson and found it to be in good agreement with the LHC results. After ensuring the consistency of the method we have examined various sparticle masses and CMSSM parameter space by imposing the various experimental constraints including the bounds from LEP data, EWPOs, B-Physics and relic density of neutralino dark matter. These results have been summarized as Figures~\ref{fig:MH1}--\ref{fig:IN}. Our analysis predict that while the neutralino LSP and lighter chargino should lie around 1.92 TeV and 2.1 TeV, gluino is expected to be found at around 7.44 TeV. 

Similarly in the scalar sector sfermion masses are found to be ranging between 5.17 TeV to 8.97 TeV. The corresponding values of CMSSM parameters $m_{0}$, $m_{1/2}$, $ A_0$ and $tan\beta$ are found to be 5.99 TeV, 3.58 TeV, $-$6.92 TeV and 36.8 respectively at the maximum marginalized entropy with three different combinational constrained space. A detailed account of the most preferred masses of various sparticle and CMSSM input parameters as found in our study is presented in Table~\ref{tab:table3}.
Clearly if the MEP technique holds good, the values in Table~\ref{tab:table3} should be most probable in experiments. However so far there 
has been no signature of supersymmetric particles at the LHC which suggests perhaps the low energy SUSY is absent. Furthermore this also suggests that it will be worthwhile to carry out the investigation on other variants of supersymmetry using the MEP.

\section*{Acknowledgements}
This work was supported in part by University Grant Commission under a Start-Up Grant no. F30-377/2017 (BSR). We thank Ravindra Yadav for his assistance regarding high-performance computing, Prabhat Solanki, Manjari Sharma and Apurba Tiwari for some valuable discussions.
We acknowledge the use of cluster computing facility at the ReCAPP, HRI, Allahabad, India during the initial phase of the work.

\end{document}